\documentclass[conference,a4paper]{IEEEtran}
\usepackage{xcolor}
\usepackage{float}
\usepackage{array}
\usepackage{bm}
\usepackage{fancyhdr}
\usepackage{enumerate}
\usepackage{amsmath}
\usepackage{graphicx}
\usepackage{booktabs}
\usepackage{amsfonts}
\usepackage{verbatim}
\usepackage{multirow}
\usepackage{graphicx, epstopdf}

\ifCLASSINFOpdf
\else
\fi
\DeclareRobustCommand*{\IEEEauthorrefmark}[1]{\raisebox{0pt}[0pt][0pt]{\textsuperscript{\footnotesize #1}}}

\begin{document}
%
\title{Channel Model in the Urban Environment for Unmanned Aerial Vehicle Communications}

\author{\IEEEauthorblockN{
Zhi Yang\IEEEauthorrefmark{1},   
Lai Zhou\IEEEauthorrefmark{2},   
Guangyue Zhao\IEEEauthorrefmark{1},    
Shidong Zhou\IEEEauthorrefmark{1}      
}                                     
\IEEEauthorblockA{\IEEEauthorrefmark{1}
Department of Electronic Engineering, Tsinghua University, Beijing, China, yang-z15@mails.tsinghua.edu.cn
}
\IEEEauthorblockA{\IEEEauthorrefmark{2}
Department of Engineering Physics, Tsinghua University,  Beijing, China}
}



\maketitle

\begin{abstract}
In order to develop and analyze reliable communications links for unmanned aerial vehicles (UAVs), accurate models for the propagation channel are required. The radio channel properties in the urban scenario are different from those in the suburb scenario and open area due to so many scattering paths from office buildings, especially when the UAV flies in the low altitude. We took some measurement campaigns on the campus of Tsinghua University with crowded apartments and office buildings. Based on the measurement result we extract the main parameters of pathloss model, and propose a simplified Saleh-Valenzuela (SV) model with specific parameters. The typical scenario of grass lawn is compared with the office buildings in the analysis of K-factor and root-mean-square (RMS) delay spread.
\end{abstract}


%
\IEEEpeerreviewmaketitle

\vspace{7pt}
\section{Introduction}
Unmanned Aerial Vehicles (UAVs) have recently attracted much interest with their high mobility and low cost, while the UAVs have come home during the past few decades, and typical examples include surveying traffic, protecting the forests and parks, reporting the headline and transporting goods \cite{handbook}. With the various application enabled by UAVs in the future, the UAVs would change the world we live in, namely, mass market UAV scenarios \cite{change}. Compared to the vehicle-to-vehicle (V2V) communication systems, the UAVs not only need gather sensor data and share information with each other, but also need cellular networks in 3D space, so wireless communication of UAVs will play an important role in the future application \cite{magazine1}\cite{zhangwei}\cite{V2V}.

\begin{figure}[!ht]\centering
    \includegraphics[width=3in]{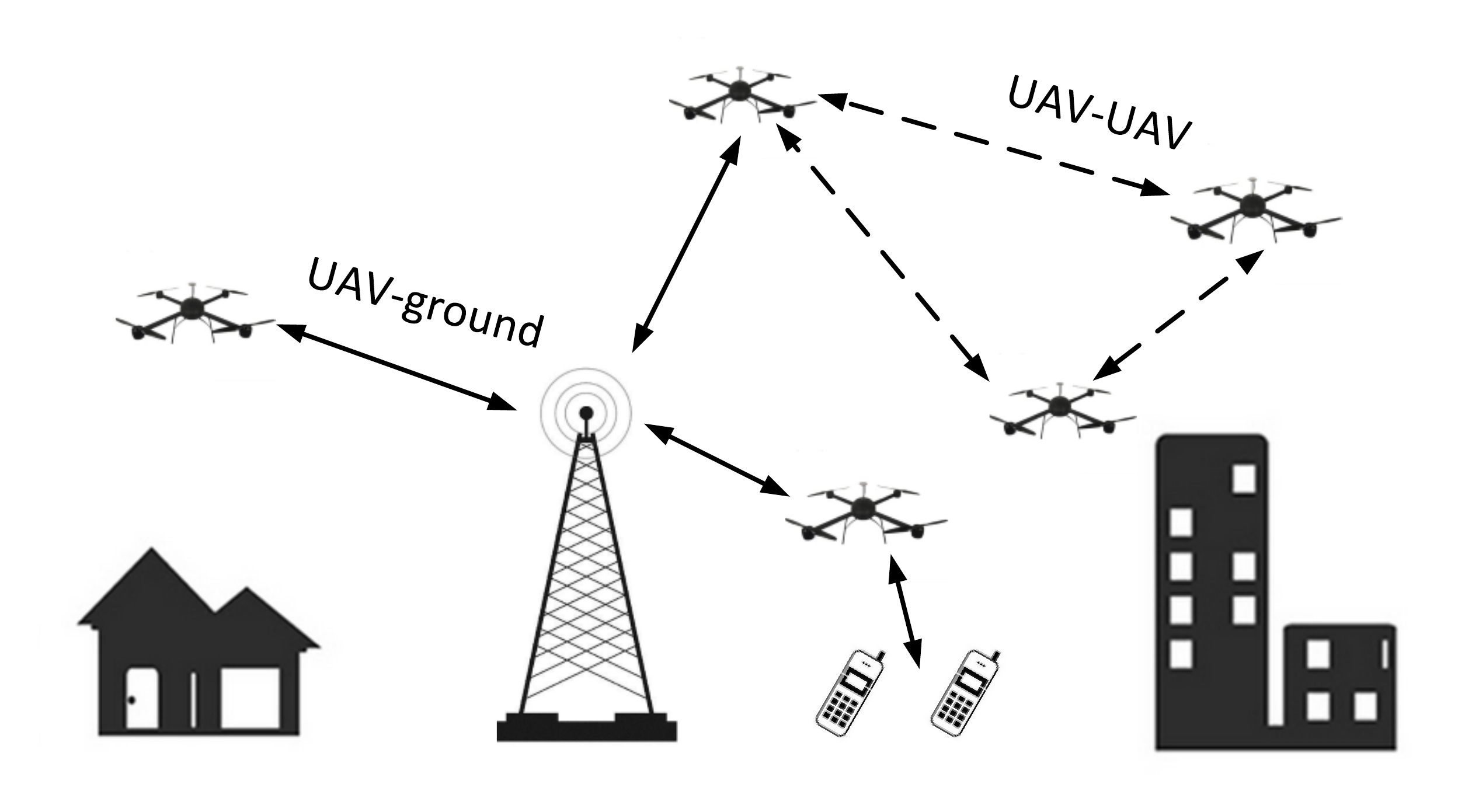}
    \vspace{-3mm}
    \caption{Different channel for UAVs communication.}
    \label{fig:fig1}
    \vspace{-3mm}
\end{figure}

As we all know, the physical wireless channel has important effect on the communication and its reliability, and using accurate channel model is critical for the system design and evaluating protocol before actual performance. In order to implement the future application of UAV cellular in the urban environment, we should have a deep understanding of the channel characteristics and study the appropriate channel model. Fig. \ref{fig:fig1} shows two types of UAVs communication links, UAV-ground and UAV-UAV channels \cite{mmwave}. The UAV-ground channels are more complicated because of complex environment and different functions, including UAV communications with base stations and cellphones.


In the past few years, much research effort has been devoted to the air-ground (AG) channels for UAVs at high-altitude \cite{survey12}, and recent applications for UAVs attract attention to low-altitude. \cite{AG3} studies the height effect on pathloss exponent (PLE) and shadow fading in the rural scenario, \cite{A2A} extends the air-to-air (AA) channel to the rice model, \cite{AG1}\cite{AG2} take measurement in the suburb environment and extract large scale parameters, including channel characteristic of multi-path components (MPCs).

Actually, physical channel study for UAVs in the urban scenario is a lack of measurement data because of difficulty in controlling the UAV through the crowded buildings, but there are many attractive applications for UAV in the city, such as delivery of goods and transportation in the air. Therefore, dedicated measurement campaigns and accurate channel models are needed for UAVs communication in the urban environment. In this article, we take some measurements in urban scenario and obtain large scale parameters in channel model, and proposes a simplified SV model with typical modeling parameters based on the result of broadband measurement. Completed channel impulse response can be reconstructed through the model in this typical environment.

The remainder of the paper is organized as follows: Section II introduces the channel measurement campaigns. In Section III, we extract the parameters of path loss model. Section IV proposes a simplified SV model. Section V compares two kinds of typical scenarios in the urban environment. Section VI concludes the paper and outlines possible future work.

\vspace{7pt}
\section{Channel Measurement campaigns}

\subsection {Measurement system}

\begin{figure}[!ht]\centering
    \includegraphics[width=3in]{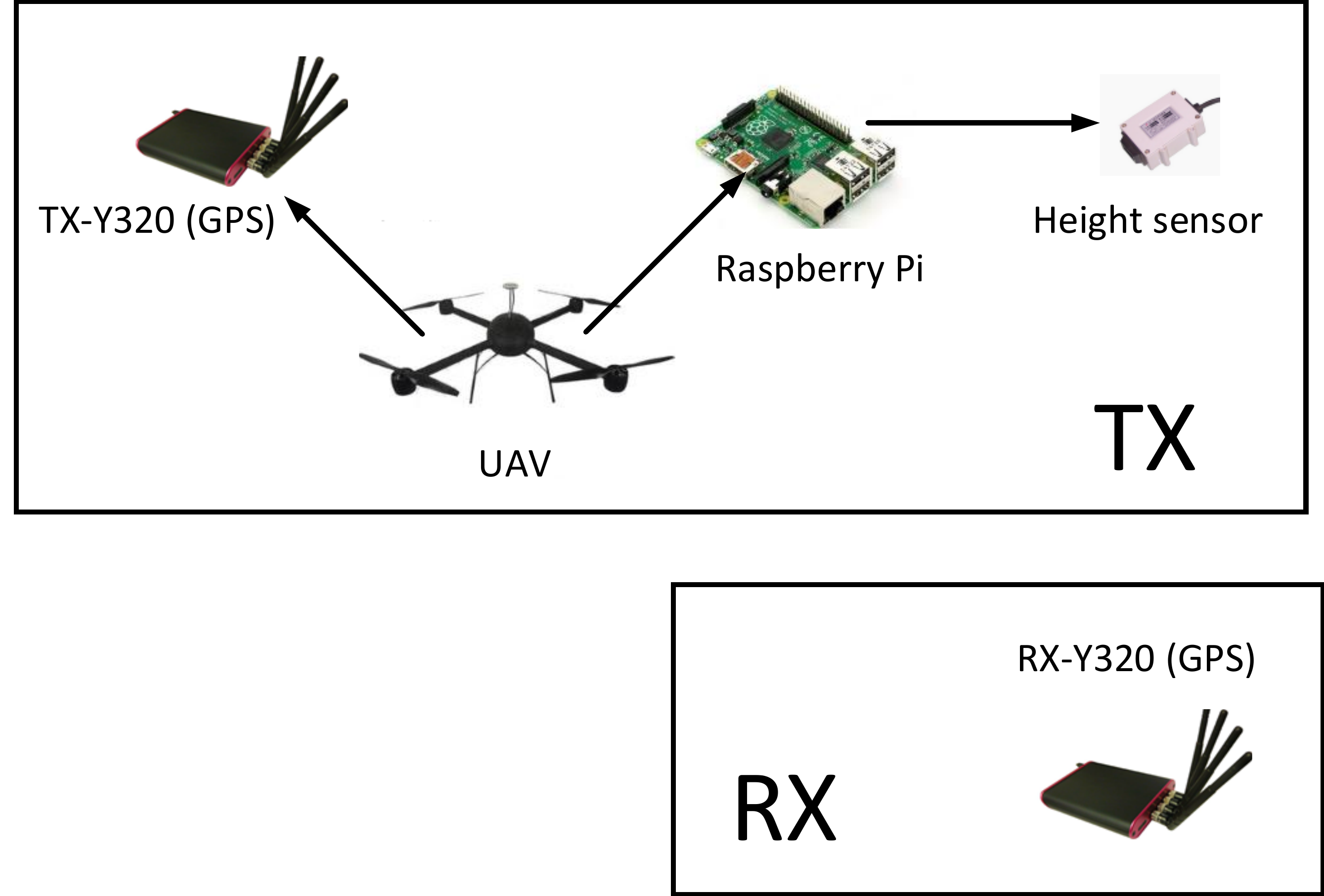}
    \vspace{-3mm}
    \caption{Measurement system.}
    \label{fig:fig4}
    \vspace{-3mm}
\end{figure}

A wide-band channel sounder with center frequency of 2.4 GHz was used to conduct channel measurement on the campus of Tsinghua University in Beijing. Fig. \ref{fig:fig4} shows the measurement system, Y320 is a kind of embedded software radio platform as transmitter (TX) and receiver (RX). The TX and RX were both equipped with GPS to record position, and the TX also carried a height sensor to record more accurate height data into Raspberry Pi. The TX was mounted on the bottom of UAV, and the RX equipped with vertically polarized and omni-directional antenna was fixed on the tripod with a height of 1 m. In order to reduce the impact from airframe fading, the dipole antenna of TX was hanged in the bottom of the UAV as shown in Fig. \ref{fig:mea}. Besides, this typical antenna gain is considered as one part of the channel in our modeling results, because of the wide use of omni-directional antenna in the UAV communications.

Details about the measurement configuration are given in Table \uppercase\expandafter{\romannumeral2}. The bandwidth is 20 MHz, and the corresponding resolution of MPC is 15 m. Considering the control security around the office buildings, the measurement range is limited to 200 m. The test signal length is 33.3 $\mu$s, and the corresponding distance is 10 km, which satisfies the maximal distance of propagation path.

\begin{table}[htbp]
\renewcommand\arraystretch{1.5}
  \centering
    \caption{\label{tab1} Measurement parameters}
    \begin{tabular}{ p{3cm} p{3cm}}
    \toprule
    \textbf{Parameter} & \textbf{Setting} \\
    \midrule
    \textbf{transmitting power} & 15dBm \\
    \textbf{Central frequency} & 2.4 GHz \\
    \textbf{Bandwidth} & 20 MHz \\
    \textbf{TX height} & 5 $\sim$ 80 m \\
    \textbf{RX height} & 1 m\\
    \textbf{Test signal length} & 33.3 $\mu$s\\
    \textbf{MPC resolution} &   15 m \\
    \bottomrule
    \end{tabular}%

\end{table}%

\subsection {Measurement scenario}
\begin{figure}[!ht]\centering
    \includegraphics[width=3in]{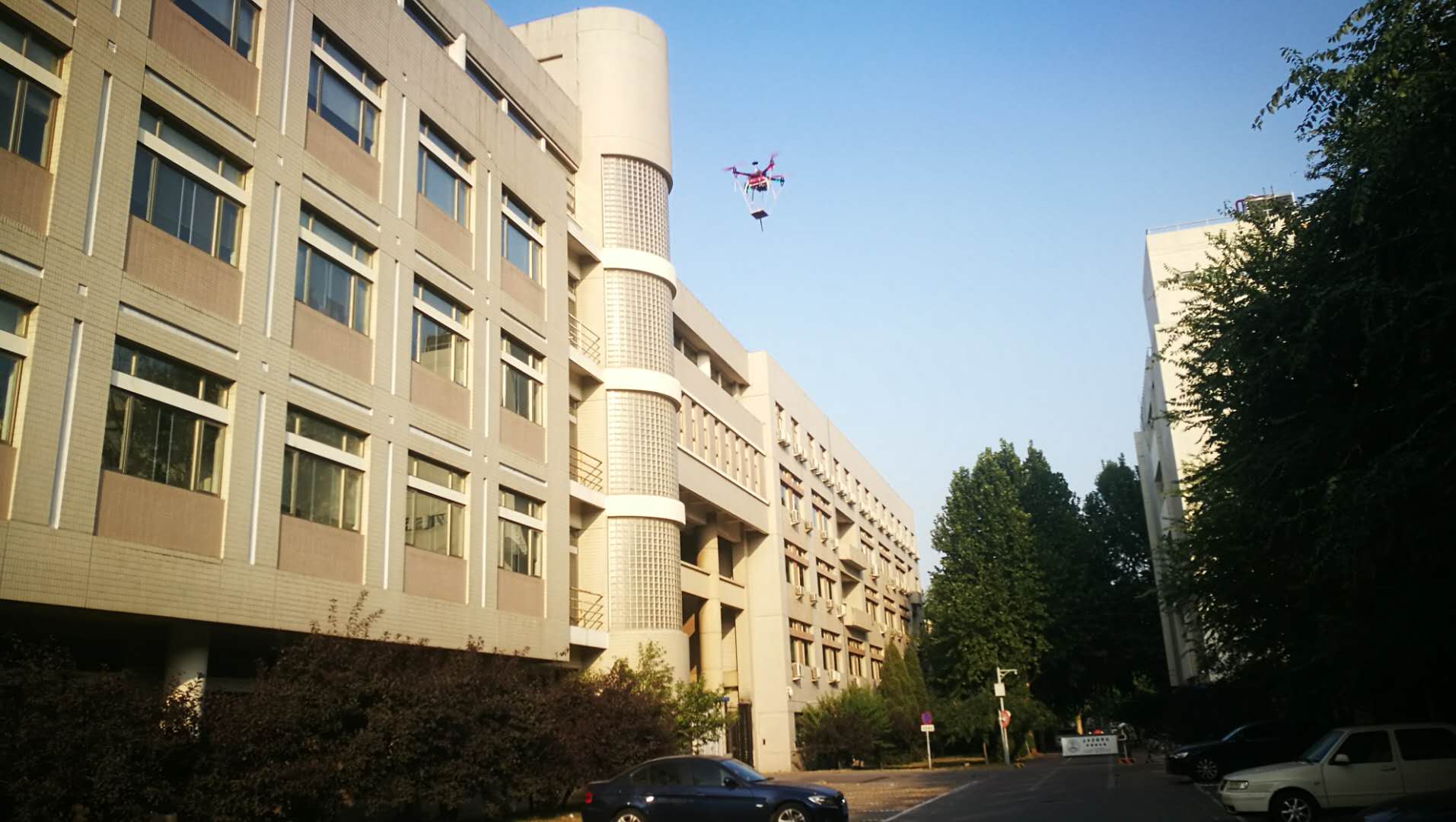}
    \vspace{-3mm}
    \caption{Measurement scenario 1 : office buildings.}
    \label{fig:fig3}
    \vspace{-3mm}
\end{figure}

\begin{figure}[!ht]\centering
    \includegraphics[width=3in]{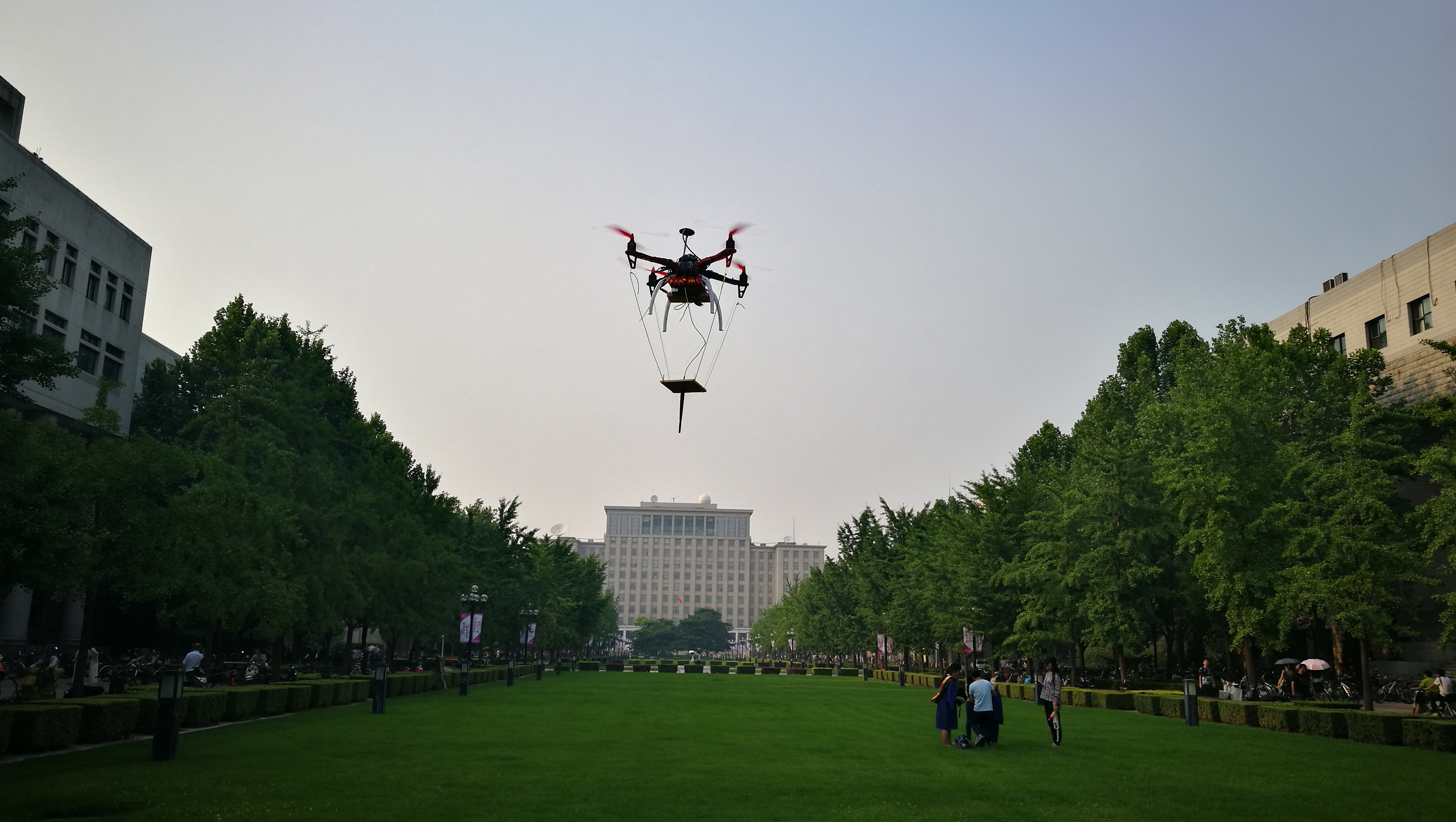}
    \vspace{-3mm}
    \caption{Measurement scenario 2 : grass lawn.}
    \label{fig:mea}
    \vspace{-3mm}
\end{figure}

The channel impulse response of each position was recorded during propagation measurement campaigns in LOS situation. Fig. \ref{fig:fig3} shows the measurement scenario around buildings, where the height of office building is about 20 m and there are some trees beside the building. In order to increase the measurement range, we took the measurement campaigns in the grass lawn as shown in Fig. \ref{fig:mea}, and the surrounding trees and buildings make it similar to the urban environment.

The UAV was controlled to fly through the streets and lawns, to collect data in different heights. Apart from a direct path between TX and RX, there should be a large number of scattering paths from surrounding trees and buildings. The receiver was placed beside the buildings or in the center of the grass lawn, which represents the green space in the urban environment.

\vspace{7pt}
\section{Path loss model}

The Subspace Alternating Generalized Expectation maximization (SAGE) algorithm \cite{SAGE} is used to obtain estimates of the MPCs from the measurement data. Based on the vertical or horizontal flight in the height of 5 m $\sim$ 80 m, we collect a large number of data in different distances, to make a omni-directional path loss model as below,

\begin{equation} \label{eq:LOSamp3}
PL(d)=PL_0+10n\cdot\log_{10}{(\frac{d}{d_0})}+S,
\vspace{0mm}
\end{equation}
where $n$ is the PLE, $d$ is the distance between TX and RX, $PL_0$ is the path loss at a reference distance $d_0$ (1 m), $S$ is the shadow fading in lognormal distribution and $S\sim N(0,\sigma^2)$.

\begin{figure}[!ht]\centering
    \includegraphics[width=3.5in]{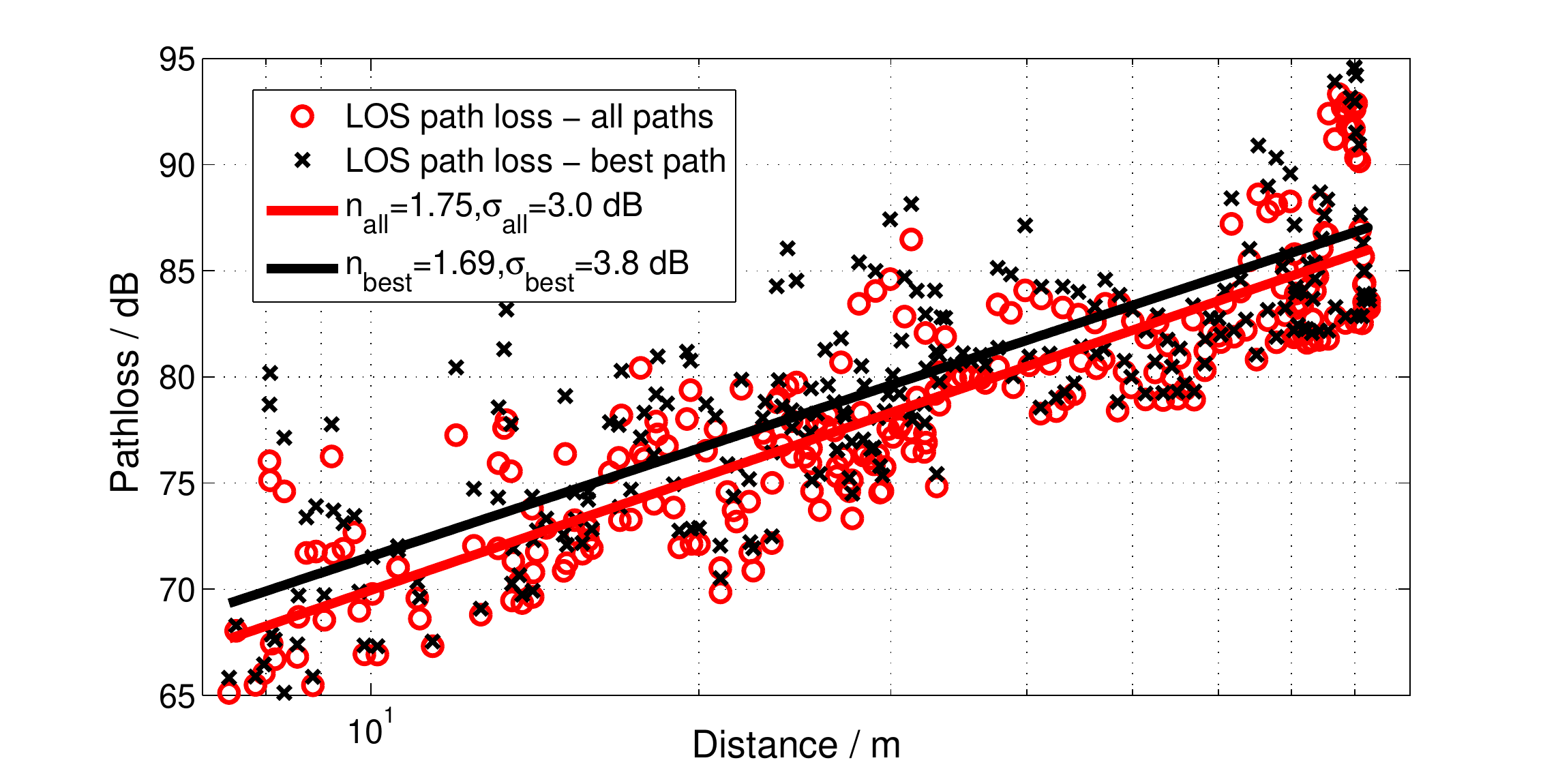}
    \caption{Path loss in the LOS scenario.}
    \label{fig:power}
\end{figure}

Fig. \ref{fig:power} shows the path loss of all paths as red circle, and best path, the most powerful path in all paths, as black cross in the line-of-sight (LOS) scenario. The PLE of all paths is 1.75 and the standard deviation is 3.0 dB. Due to the direct path dominating the channel impulse response in the LOS scenario, the fitting parameters of best path are similar to that of all paths. There are many scattering paths from surrounding trees and buildings, besides, the vertical angle between the TX and RX continuously decreases when the UAV moves away from the the receiver, leading to the increase of antenna gain in LOS path loss. Therefore, the PLE of all path and best path are both less than 2.

\begin{figure}[!ht]\centering
    \includegraphics[width=3.5in]{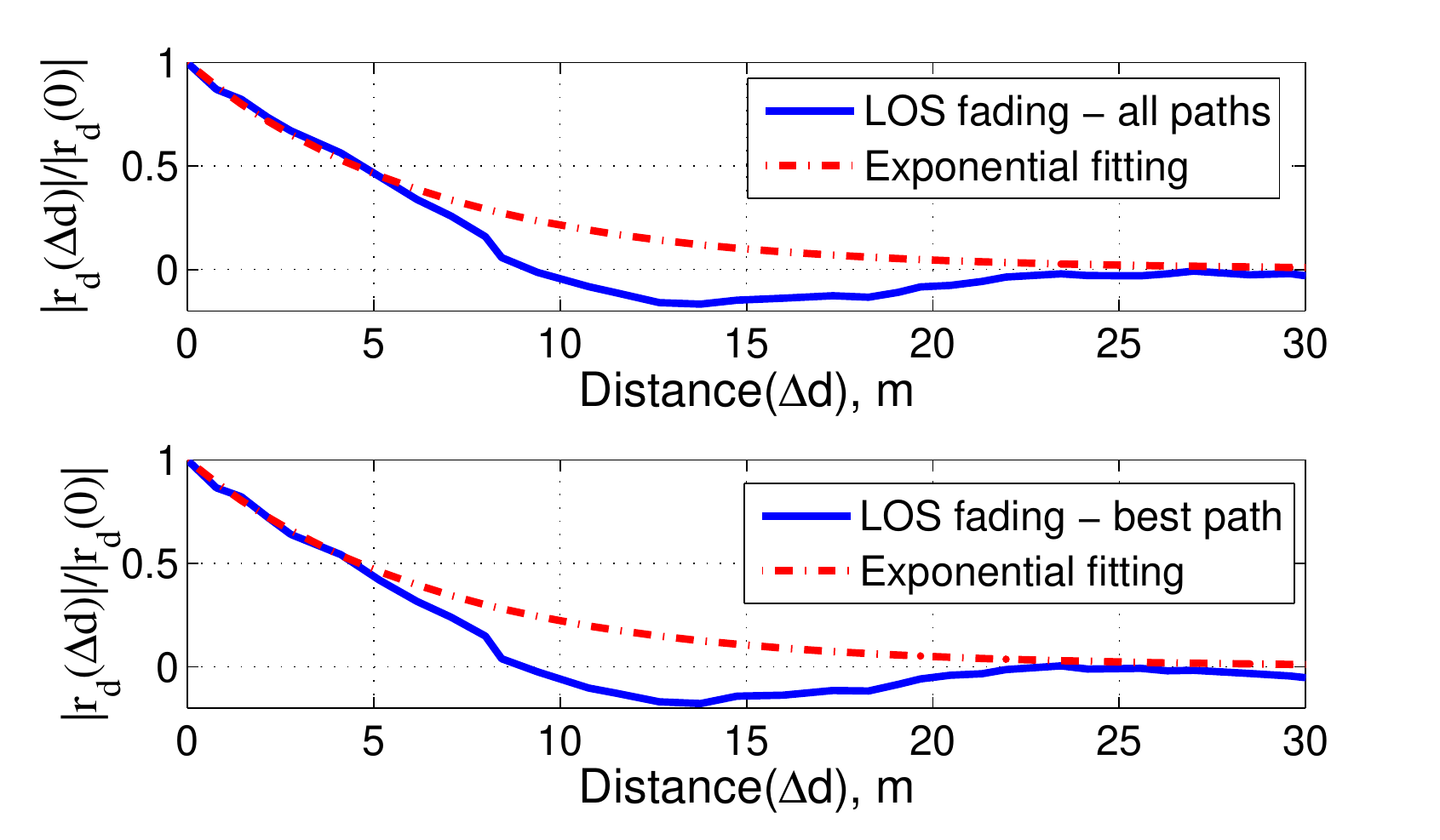}
    \vspace{-3mm}
    \caption{Distance based autocorrelation for shadow fading.}
    \label{fig:corr}
    \vspace{-3mm}
\end{figure}

The distance dependent autocorrelation function of shadow fading could be calculated as
\begin{equation}
  r_d(\Delta d) = E[S(d)S(d+\Delta d)],
\end{equation}
where $S(d)$ represent the shadow fading in the distance of $d$, and $\Delta d$ is the distance between two positions. The exponential auto-correlation function \cite{corr} is commonly used to describe the large-scale fading as
\begin{equation}
  r_d(\Delta d) = \sigma^2e^{-\frac{ln2}{d_0}(\Delta d)},
\end{equation}
where $d_0$ is the de-correlation distance, and the fitting results are 4.5 m and 4.6 m for all paths and best path, respectively. And the typical de-correlation distance of cellular network in the urban environment is also about 5 m \cite{corr}. Fig. \ref{fig:corr} shows the normalized autocorrelation function of LOS fading and exponential fitting line.

\vspace{7pt}
\section{Saleh-Valenzuela model}

The SV channel model \cite{SV} is usually used to model the measured results in the indoor environment because of abundant scattering paths from walls and furniture. Based on our measurement we find that the channel impulse response in the urban environment could be modeled as a simplified S-V model. Assuming that the arriving time of the most powerful path in each snapshot is 0, the model could be described as:

\begin{equation}
  h(t) = a_0+\sum_{i=1}^{N_f} a_{f,i} \delta (t-\tau_{f,i})+\sum_{i=1}^{N_b} a_{b,i} \delta (t-\tau_{b,i}).
\end{equation}
The conventional model is characterized by inter-cluster and intra-cluster parameters, but the simplified S-V model only has one cluster containing three parts, the central ray with amplitude $a_0$ and delay 0, the $i$-th pre-cursor ray with amplitude $a_{f,i}$ and delay $\tau_{f,i}$, and the $i$-th post-cursor ray with amplitude $a_{b,i}$ and delay $\tau_{b,i}$. $N_f$ and $N_b$ are the numbers of pre-cursor rays and post-cursor rays, respectively.

\begin{figure}[!ht]\centering
    \includegraphics[width=3.5in]{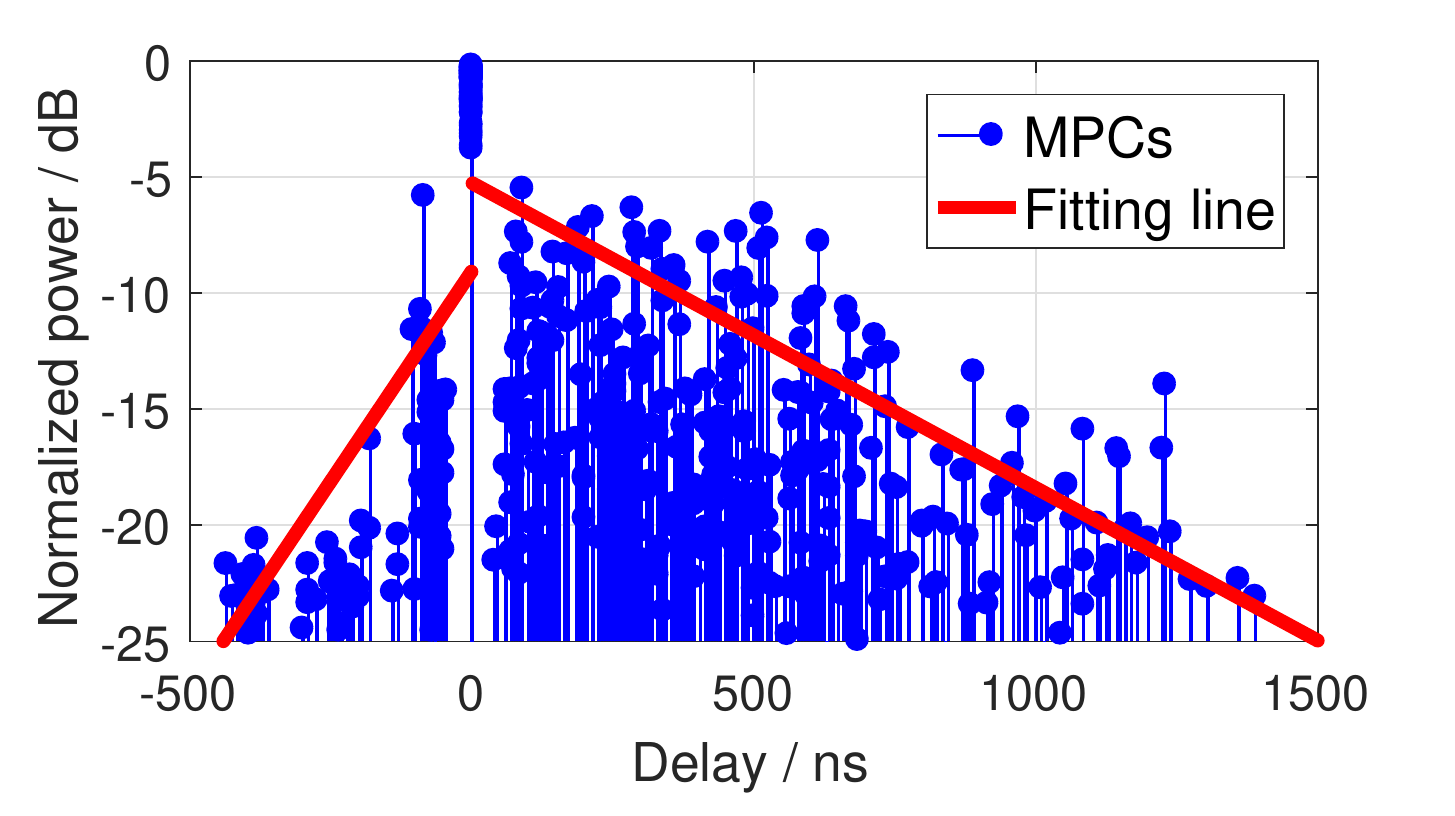}
    \vspace{-3mm}
    \caption{Estimated MPCs with SAGE algorithm and Fitting results in scenario 2.}
    \label{fig:sv}
    \vspace{-3mm}
\end{figure}

The average amplitudes of $a_{f}$ and $a_{b}$ could be modeled as exponential decay with power decay times $\gamma_{f}$ and $\gamma_{b}$, 
\begin{equation}
    \begin{aligned}
  a_{f}(\tau) = a_f(0)e^{-|\tau| / \gamma_{f}}\\
    a_{b}(\tau) = a_b(0)e^{-|\tau| / \gamma_{b}},
    \end{aligned}
\end{equation}
where the amplitude bases of $a_{f}(0)$ and $a_{b}(0)$ are coupled with the amplitude of the central ray $a_{0}$ by K-factors that are defined as,
\begin{equation}
    \begin{aligned}
  K_{f}(\tau) = 20log_{10}(\left| \frac{a_0}{a_f(0)} \right|)\\
    K_{b}(\tau) = 20log_{10}(\left| \frac{a_0}{a_b(0)} \right|).
    \end{aligned}
\end{equation}
Fig. \ref{fig:sv} merges the normalized power of MPCs from multiple snapshots together and shows the fitting lines in the time domain.
In most cases, the most powerful path is the direct path. However, when there are more antenna gain in the scattering paths, these paths could be more stronger than the direct path and the delay of the strongest path is defined as 0 in (4), resulting in some paths with delays less than 0 in the plot.

What is more, there are no MPCs with delays near 0, since the most powerful path dominates the channel and the nearby paths are merged together in our estimated method, so that the offset time should be considered when modeling the ray arrival time.

\begin{figure}[!ht]\centering
    \includegraphics[width=3.5in]{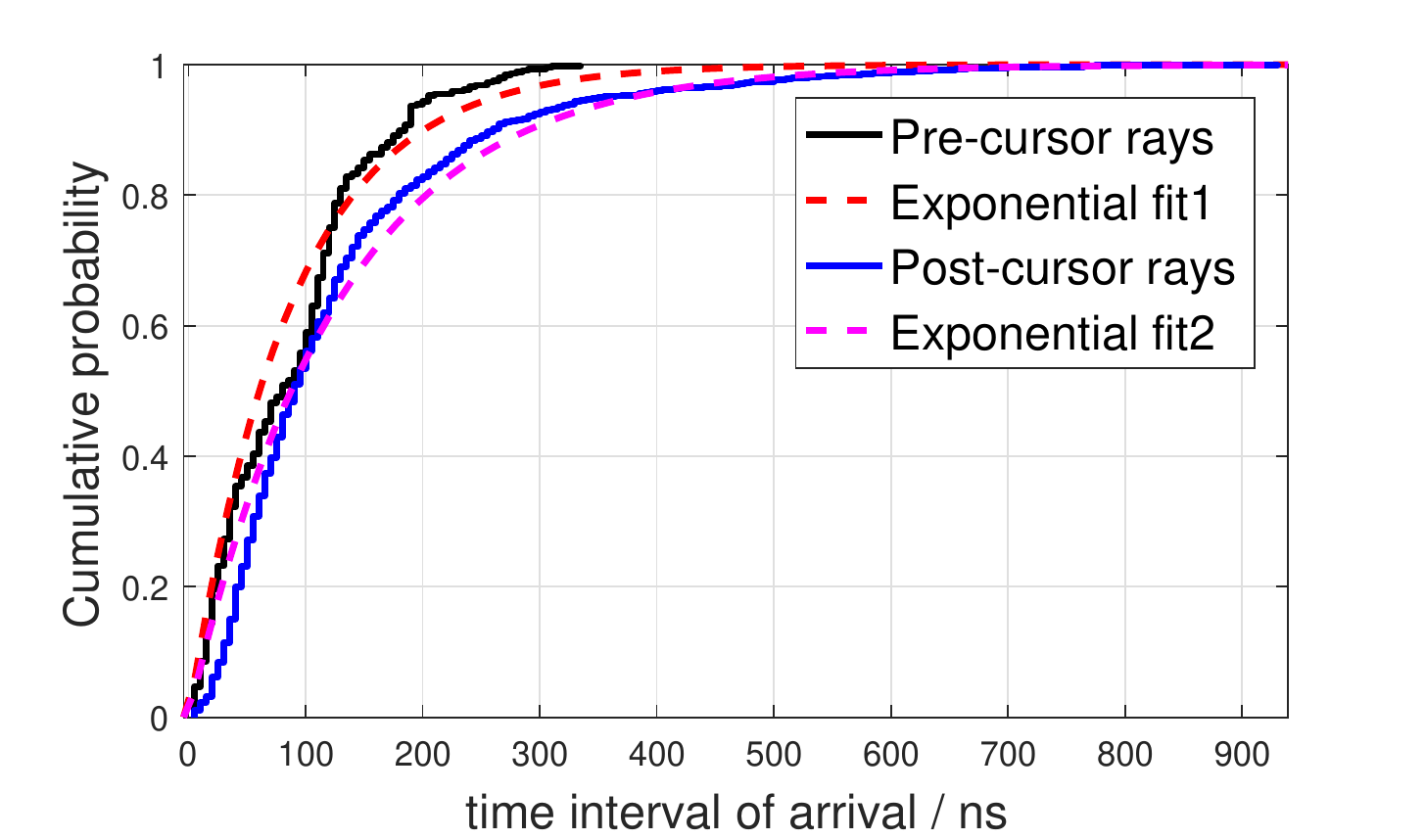}
    \vspace{-3mm}
    \caption{Time interval of ray arrival and Fitting results in scenario 2.}
    \label{fig:possion}
    \vspace{-3mm}
\end{figure}

The distributions of the ray arrival times are given by the Poisson processes as
\begin{equation}
    \begin{aligned}
  p(\tau_{f,i}|\tau_{f,i-1}) = \lambda_{f} exp[-\lambda_{f}(\tau_{f,i}-\tau_{f,i-1})]\\
    p(\tau_{b,i}|\tau_{b,i-1}) = \lambda_{b} exp[-\lambda_{b}(\tau_{b,i}-\tau_{b,i-1})].
    \end{aligned}
\end{equation}
where $\lambda_{f}$ and $\lambda_{b}$ are the arrival rates of pre-cursor and post-cursor rays, respectively. Fig. \ref{fig:possion} shows the cumulative distribution function (CDF) plot of the time interval of the arrival rays and the measured results match the exponential fitting well. Hence, the ray arrival times could be expressed as the Poisson process, and the offset time is 50 $ns$. The parameters of SV model are summarized in Table II. 

\begin{table}[htbp]
\renewcommand\arraystretch{1.5}
  \centering
    \caption{\label{tab sv} SV model parameters}
    \begin{tabular}{ p{4.5cm} p{0.8cm} p{0.8cm} p{0.8cm}}
    \toprule
    \textbf{Parameters}           & \textbf{Notation}     & \textbf{scenario 1}     & \textbf{scenario 2}  \\
    \midrule
    Pre-cursor rays K-factor / dB         & $K_f$           &  8.1           &  11.4\\
    Pre-cursor rays power decay time / ns   &$\gamma_f$         &   240           &   316\\
  Pre-cursor arrival rate / $ns^{-1}$       &$\lambda_f$        & 0.0092    &0.0075  \\
    Number of pre-cursor rays         &$N_f$            &  2.2               &   1.6\\
    Post-cursor rays K-factor / dB          &$K_b$            &  2.8             & 5.1  \\
    Post-cursor rays power decay time / ns    &$\gamma_b$         &    448           &  662 \\
  Post-cursor arrival rate / $ns^{-1}$          &$\lambda_b$        & 0.0073   &  0.0057 \\
    Number of post-cursor rays        & $N_b$           &    4.8              &  5.4\\
    \bottomrule
    \end{tabular}%
\end{table}%

\vspace{7pt}
\section{K factor and RMS Delay Spread}
\begin{figure}[!ht]\centering
    \includegraphics[width=3.5in]{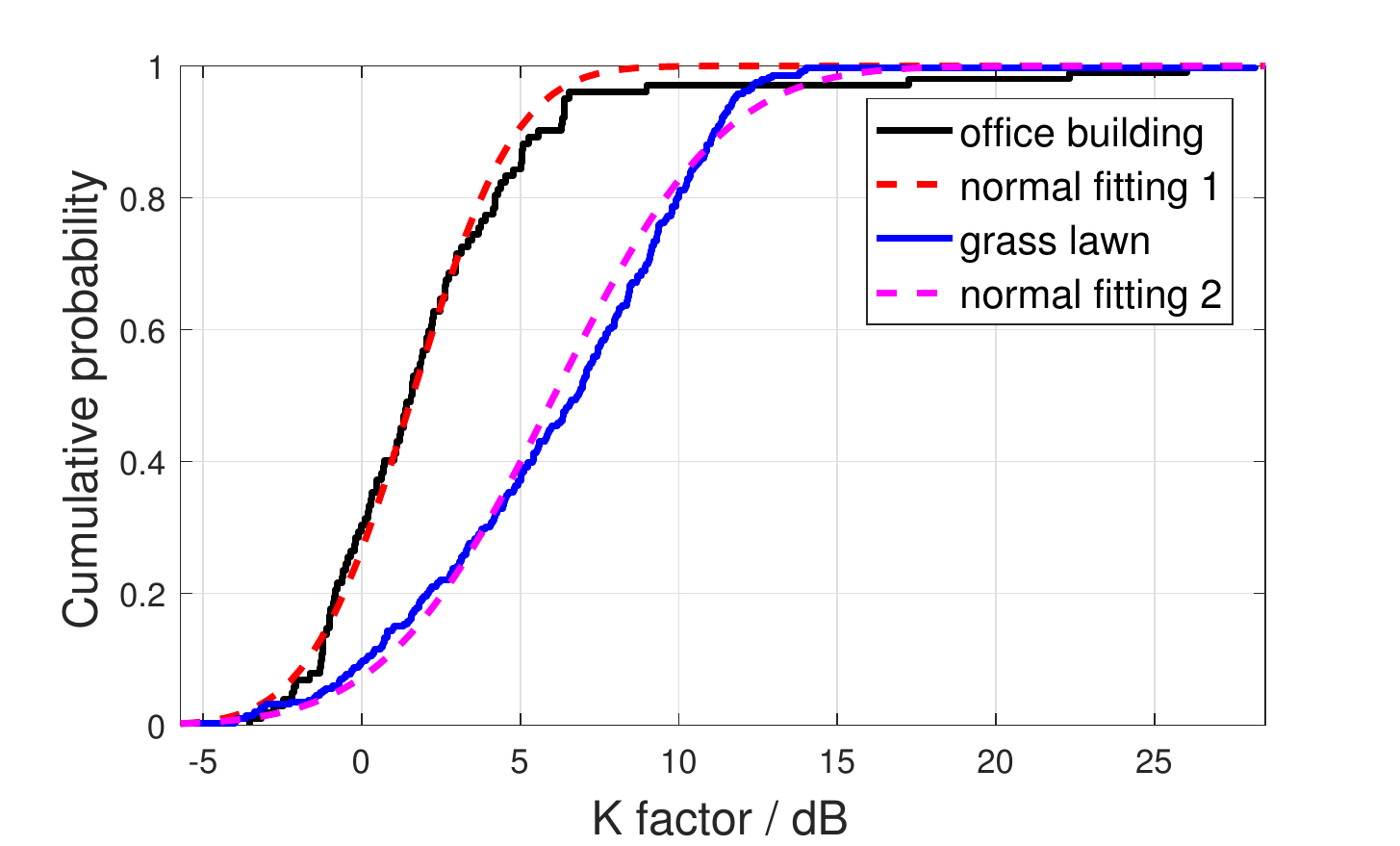}
    \vspace{-3mm}
    \caption{Measurement results of K factor}
    \label{fig:kfactor}
    \vspace{-3mm}
\end{figure}

In order to have a deep understanding of the typical scenario in the urban environment, i.e., the office buildings and grass lawn, the K-factor and RMS delay spread are selected to express the channel characteristics. K-factor is the important parameter to model the small scale fading as

\begin{equation} \label{eq:kfactor}
K=10log_{10}(\frac{P_{LOS}}{\sum_{i=1}^{N-1} P_{NLOS,i}}),
\vspace{0mm}
\end{equation}
where $P_{LOS}$ is the power of direct path, $P_{NLOS,i}$ is the power of $i^{th}$ non-line-of-sight (NLOS) path, $i=1,...,N-1$, and $N$ is the number of the MPCs. Fig. \ref{fig:kfactor} shows the CDF plot of the K-factor, and the measurement result of grass lawn is larger than that of office buildings, because the reflection paths from nearby buildings are stronger than the scattering paths from trees.

\begin{figure}[!ht]\centering
    \includegraphics[width=3.5in]{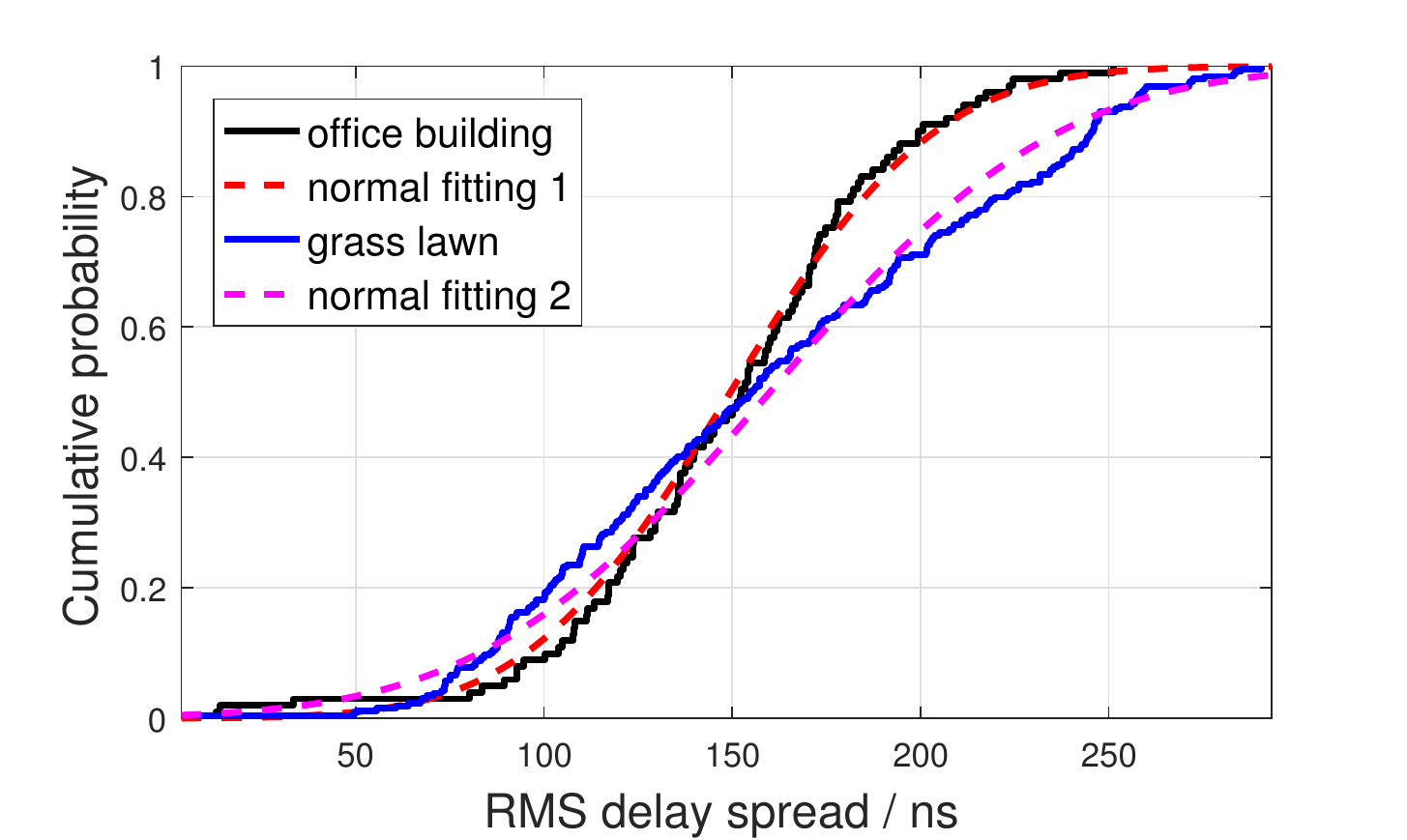}
    \vspace{-3mm}
    \caption{Measurement results of RMS delay spread}
    \label{fig:rms}
    \vspace{-3mm}
\end{figure}

Furthermore, considering the importance of designing pilot patterns and determining achievable frequency diversity, the RMS delay spread are calculated as

\begin{equation} \label{eq:k-factor}
\bar{\tau}=\frac{\sum_{i=1}^{N} P_i \cdot \tau_i}{\sum_{i=1}^{N} P_i},
\vspace{0mm}
\end{equation}

\begin{equation} \label{eq:rmsds}
\tau_\sigma=\sqrt{\frac{\sum_{i=1}^{N} P_i(\tau_i-\bar{\tau})^2}{\sum_{i=1}^{N} P_i}},
\vspace{0mm}
\end{equation}
where $P_i$ is the power of $i^{th}$ MPC with a delay $\tau_i$, and $\bar{\tau}$ is the weighted mean value of the delay. The measured data of RMS delay spread in two kinds of scenarios are shown in Fig. \ref{fig:rms}, and the result of grass lawn is larger than that of office buildings due to larger propagation range. The normal distribution performs better than the lognormal distribution when fitting the results, for example, the value of log likelihood is -521 in normal fitting compared with -558 in lognormal fitting in the office building scenario. Hence, the normal fitting is used to model the measurement data and the Table III summarizes the fitting parameters as below.

\begin{table}[htbp]
\renewcommand\arraystretch{1.5}
  \centering
    \caption{\label{tab3} Measurement results}
    \begin{tabular}{ p{2.3cm} p{1.2cm} p{1cm} p{1cm}}
    \toprule
    \textbf{Parameters} & \textbf{Notation} & \textbf{scenario 1} & \textbf{scenario 2} \\
    \midrule
    \multirow{2}{2.3cm}{K factor} & $\mu_k$ / dB & 2.20 & 6.15 \\
             & $\sigma_k$ / dB & 4.32& 4.36\\
    \multirow{2}{2.3cm}{RMS delay spread} & $\mu_{\tau}$ / ns &149.50 & 159.97 \\
                     & $\sigma_{\tau}$ / ns  &42.44 & 60.15\\
    \bottomrule
    \end{tabular}%
\end{table}%

\vspace{7pt}
\section{Conclusion}

Based on the measurement data in the urban environment, this article provides the parameters of path loss model suitable for long time prediction. Although there are some scattering paths, the direct path still dominates the channel. A simplified SV model is proposed and the modeling parameters of two kinds of scenarios are summarized. With the above two models, the completed channel impulse response is obtained to evaluate the communication system of UAV in this typical urban environment. What is more, the K factor and RMS delay spread are calculated to make a comparison between the office buildings and grass lawn, so the difference between different scenarios should be taken into consideration.

Considering the difficulty in taking such measurement in the urban environment, the recorded data is limited in this paper and more measurement should be carried out to propose a more accurate and general model. For example, due to the varied building density in height, the parameters such as PLE would be impacted because the UAV will change its height flexibly. The AG channel should be extended to the AA channel when there are a large number of UAVs communication in the air. Furthermore, the channel sounder should be capable of a larger bandwidth and better sensitivity, and the effect of antenna pattern and flight attitude need to be studied.

\section{acknowledgment}
The research presented in this paper has been kindly funded by the projects as follows, National S\&T Major Project (2017ZX03001011), National Natural Science Foundation of China (61631013), Foundation for Innovative Research Groups of the National Natural Science Foundation of China (61621091),  Tsinghua-Qualcomm Joint Project, Future Mobile Communication Network Infrastructure Virtualization and Cloud Platform (2016ZH02-3) .


\end{document}